\documentclass[fleqn,usenatbib]{mnras}

\usepackage{newtxtext,newtxmath}

\usepackage[T1]{fontenc}
\usepackage{ae,aecompl}

\usepackage{graphicx}	
\usepackage{amsmath}	
\usepackage{amssymb}	
\usepackage{booktabs}
\usepackage{multirow}
\usepackage{comment}

\newcommand{\gaia}{{\it Gaia}}

\newcommand{\NGTS}{{\it NGTS}}

\newcommand{\Nstar}{NGTS J052218.2-250710.4}

\newcommand{\NstarMassA}{$0.17391 ^{+0.00153}_{-0.00099}$} 
\newcommand{\NstarMassB}{$0.17418 ^{+0.00193}_{-0.00059}$} 
\newcommand{\NstarRadA}{$0.2045 ^{+0.0038}_{-0.0058}$} 
\newcommand{\NstarRadB}{$0.2168 ^{+0.0047}_{-0.0048}$} 
\newcommand{\SemiMajorAxis}{$4.2945 ^{+0.0158}_{-0.0039}$}
\newcommand{\OrbitalInc}{$87.404 ^{+0.016}_{-0.019}$}

\newcommand{\Nperiod}{$1.747752$}
\newcommand{\Nduration}{$1.3458$}

\newcommand{\kms}{km\,s$^{-1}$}

\newcommand{\masy}{mas\,yr$^{-1}$}

\newcommand{\cmss}{cm\,s$^{-2}$}

\newcommand{\LSO}{La Silla Observatory}

\title[NGTS discovery of an M dwarf binary]{A low-mass eclipsing binary within the fully convective zone from the NGTS}

\author[S.L.Casewell et al., ]{
\parbox{\textwidth}{
S. L. Casewell$^{1}$\thanks{E-mail: \href{slc25@le.ac.uk}{slc25@le.ac.uk}}, 
L. Raynard$^{1}$,
C. A. Watson$^{2}$,
E. Gillen$^{3,\dagger}$,
E. de Mooij$^{2}$,
D. Bayliss$^{4,5}$,
F. Bouchy$^{5}$,
A. Thompson$^{2}$,
J.~A.~G. Jackman$^{4}$,
M. R. Burleigh$^{1}$,
A. Chaushev$^{1}$,
C. Belardi$^{1}$,
T. Louden$^{4}$,
M.R.Goad$^{1}$,
L.~D.~Nielsen,$^{5}$
K. Poppenhaeger$^{2}$,
Ph. Eigm\"uller$^{6}$,
Maximilian~N.~G{\"u}nther$^{3}$,
J. S. Jenkins$^{7}$,
J. McCormac$^{4}$,
M. Moyano$^{8}$,
D. Queloz$^{3}$,
A.~M.~S.~Smith$^{6}$,
R. G. West$^{4}$,
P. J. Wheatley$^{4}$
}
\\
$^{1}$Dept.\ of Physics and Astronomy, Leicester Institute of Space and Earth Observation, University of Leicester, University Road,\\ Leicester, LE1 7RH, UK\\
$^{2}$Astrophysics Research Centre, School of Mathematics and Physics, Queen's University Belfast, BT7 1NN, Belfast, UK\\
$^{3}$Astrophysics Group, Cavendish Laboratory, J.J. Thomson Avenue, Cambridge CB3 0HE, UK\\ 
$^{\dagger}$ Winton Fellow \\
$^{4}$Dept.\ of Physics, University of Warwick, Gibbet Hill Road, Coventry CV4 7AL, UK\\
$^{5}$Observatoire de Gen{\`e}ve, Universit{\'e} de Gen{\`e}ve, 51 Ch. des Maillettes, 1290 Sauverny, Switzerland\\
$^{6}$Institute of Planetary Research, German Aerospace Center, Rutherfordstrasse 2, 12489 Berlin, Germany\\
$^{7}$Departamento de Astronom\'ia, Universidad de Chile, Camino El Observatorio 1515, Las Condes, Santiago, Chile\\
$^{8}$Instituto de Astronom\'ia, Universidad Cat\'olica del Norte,
      Av. Angamos 0610, Antofagasta, Chile.
}

\date{Accepted XXX. Received YYY; in original form ZZZ}

\pubyear{2017}

\begin{document}
\label{firstpage}
\pagerange{\pageref{firstpage}--\pageref{lastpage}}
\maketitle

\begin{abstract}
We have discovered a new, near-equal mass, eclipsing M dwarf binary from the Next Generation Transit Survey. This system is only one of 3 field age ($>$ 1 Gyr), late M dwarf eclipsing binaries known, and has a period of 1.74774 days, similar to that of CM~Dra and KOI126. Modelling of the eclipses and radial velocities shows that the component masses are $M_{\rm pri}$=\NstarMassA $M_{\odot}$, $M_{\rm sec}$=\NstarMassB $M_{\odot}$; radii are $R_{\rm pri}$=\NstarRadA $R_{\odot}$, $R_{\rm sec}$=\NstarRadB. 
The effective temperatures are $T_{\rm pri} = 2995\,^{+85}_{-105}$\,K and $T_{\rm sec} = 2997\,^{+66}_{-101}$\,K, consistent with M5 dwarfs and broadly consistent with main sequence models.
This pair represents a valuable addition which can be used to constrain the mass-radius relation at the low mass end of the stellar sequence.

\end{abstract}




\section{Introduction}

M dwarfs are the most numerous stellar spectral type in the Galaxy \citep{henry97}, and have recently become the focus of many transiting planet searches (e.g., TRAPPIST (\citealt{jehin11}), the Next Generation Transit Survey (NGTS: \citealt{ngts}), MEarth (\citealt{berta12})) since their small radii make it easier to detect Earth sized planets.  In fact, it appears M dwarfs are more likely to host planets than  higher mass stars, and the majority of these are Earth-like, as there is a known dearth of  hot Jupiters around  M dwarfs, with only three known to date: NGTS-1b \citep{bayliss17}, Kepler-45b \citep{johnson12} and HATS-6b \citep{hartman15}.

It is possible to directly measure the radius of M dwarfs via interferometry (e.g. \citealt{Demory2009,Boyajian2012}), however, there are currently no interferometric measurements for M dwarfs later than M6.  For the majority of isolated M dwarfs, the mass and radius can only be determined by using evolutionary models such as the BCAH15 \citep{baraffe15} or PARSEC models \citep{parsec}, combined with luminosity measurements, and a mass-luminosity relation (e.g. \citealt{delfosse00}). Therefore it is of great importance that these models are verified to be accurate.
\citet{torres13} highlighted that for detached, eclipsing binaries containing M dwarfs (where the radius can be directly measured) inflated radii and cooler effective temperatures in comparison with the theoretical models, are commonly reported. These two effects can roughly cancel each other out, giving a luminosity that is consistent with the model, hinting that this  phenomenon is a located on the surface of the star (for instance, due to activity e.g., \citealt{Stelzer2013}, and metallicity e.g., \citealt{stassun2012,lopez07}). For most exoplanet hosts there is no direct measurement of mass and radius, evolutionary models are used and hence this effect will lower the precision of the  radii and mass (hence the density) of any planet present. 

A study by \citet{brown11} showed that while the $Kepler$ KIC parameters are accurate for sun-like stars, they are less reliable for T$_{\rm eff}<$3750 K, where the M dwarfs are fully convective. \citet{muirhead12} also find that the radii for KIC M dwarfs are generally underestimated  in the KIC \citep{borucki11}, as do \citet{gaidos12} in their comparison between the KIC and M2K M dwarf survey \citep{apps10}. \citet{mann17} obtained improved parallaxes for eight KIC systems containing M dwarf primary stars, including $Kepler-42$ (KOI 961: \citealt{murihead12a}), which is fully convective. They concluded that once they accounted for the eccentricity of the orbits for the majority of systems, the M dwarfs had masses and radii consistent with the stellar models, however, $Kepler-42$ was still an outlier. Model based densities gave a radius $\sim$6 per cent smaller than empirical determinations from the luminosity, and consequently an effective temperature too hot by $\sim$2 per cent. The $Kepler$ lightcurve shows that the star's variability is less than one per cent, indicating that in this case, activity is unlikely to be the cause of the discrepancy in the radius determinations. Another study by \citet{kesseli} used a combination of mass-radius relations and vsini measurements for the MEarth M dwarfs. They find that stellar evolutionary models underestimate the radii by 10-15 per cent, but that at higher masses (M $>$ 0.18 M$_{\odot}$) the discrepancy is only about 6 per cent, comparable to the results from interferometry and eclipsing binaries. At the lowest masses (0.08 $<$ M $<$ 0.1  M$_{\odot}$), they find the discrepancy between observations and theory is between 13 and 18 per cent, and unlikely to be due to effects from age and rotation.   \citet{Mann15} derived an empirical relationship between T$_{\rm eff}$ and radius using spectra of 187 K7-M7 dwarfs, They also found their best fit models overpredicted effective temperature by 2.2 per cent and underpredict radii by 4.6 per cent. They suggest this difference is more likely to be related to the models (e.g. mixing length, opacities) than stellar parameters such as metallicity, activity or rotation. 

In the rare cases where direct measurements of masses and radii can be made (i.e. low-mass stars in eclipsing binary systems), the radii determined are also often inconsistent with evolutionary models, by as much as 10 per cent \citep{feiden12, terrien12}. For example, \citet{birkby12} determined the radii of their eclipsing M dwarfs were inflated by as much as 3-12 per cent, and that this inflation was not dependent on the activity of the M dwarf. However, this is not a consistent deviation from models - some objects agree very well, for instance the interferometric values determined by \citet{demory09} for M0-M5.5 stars.  The models of \citet{spada13}, T$_{\rm eff}$-radius relationships of \citet{Mann15} and observations of \citet{Boyajian2012} suggest that both binary and single M dwarfs suffer from the aforementioned radius discrepancy, and that this discrepancy is largest in binaries with short periods ($<$ 1.5 days) and in lone, low mass (M$<$0.4 M$_{\odot}$), low metallicity objects. This inflation for objects in short period binaries is also seen by \citet{Boyajian2012} and \citet{kraus11} who suggest it is due to the presence of the companion star, probably because they are tidally locked into very high rotation speeds that may enhance activity and inhibit convection.  

Another issue is that some areas of parameter space surrounding the M dwarfs are only sparsely sampled. In particular the later spectral types.  For instance TRAPPIST-1, an M dwarf hosting 7 rocky planets \citep{gillon17}, has a mass of 0.082 M$_{\odot}$. There are very few model-independent mass-radius measurements in this range, one of which is CSS03170 a low-mass star in an eclipsing binary with a white dwarf \citep{parsons12}. Another system is EBLM J0555-57 \citep{boetticher17}, which is a low mass M dwarf (0.081 M$_{\odot}$, 0.084 R$_{\odot}$, log g=5.5) orbiting a sun-like star. This object which is unusually dense for an M dwarf, has a log g similar to that of many field brown dwarfs e.g. \citep{kirkpatrick99,chabrier00}. This M dwarf has masses and radii consistent with the models, and the NextGen models \citep{baraffe98} used in \citep{boetticher17} suggest that EBLM J0555-57B would have an effective temperature of 2200 K, making it an M9 dwarf or later \citep{rajpurohit13}.   There are a few other objects within this low mass regime with main sequence companions : \citet{triaud17} performed a survey for M dwarfs in binaries with FGK stars. 31 per cent of their sample had masses of less than 0.2 M$_{\odot}$. However, most of these are too faint to be directly detected in the spectra, hence these are single lined binaries and as yet, no accurate radii are available for these systems. The \gaia\ parallaxes will allow higher precision radius measurements to be made using the mass-luminosity relations for the late M dwarfs (e.g. \citealt{Mann15}). \citet{yilen} discovered 1SWASPJ011351.29+314909.7, a low mass M dwarf in a binary with a much higher mass main sequence star. This system is only a single lined binary, and no secondary eclipse is detected, yet, the mass and radius are consistent with evolutionary models. The effective temperature is $\sim$600 K hotter than the models predict. This effect is also the case for the M dwarf-main sequence binary KIC 1571511 \citep{ofir}. These two objects produce results that are conflicting with those of \citet{torres13}, who found effective temperatures that were consistently cooler than the models predicted. 

It is clear that in order to achieve a measure of accuracy in planetary mass and radius measurements, we require accurate masses and radii for the primary stars. As models of fully convective M dwarfs seem to overpredict effective temperatures and underpredict the radii (e.g. \citealt{Morales09,kraus11,kesseli}), it is important to observe as many double-lined eclipsing systems as possible to build an empirical mass-radius relation to complement the theoretical ones such as those of \citet{spada13,kraus11,torres10,torres13}.

\section{Observations}
\Nstar{} was discovered using \NGTS{} photometry in conjunction with z' band follow-up photometry from SHOC (SAAO 1m), near-IR spectra from LIRIS (WHT) and radial velocity measurements from HARPS spectra (ESO 3.6m). We detail these observations in this Section and provide a summary in Table \ref{tab:obs_summary}.

\begin{table*}
	\centering
	\caption{Summary of observations which led to the discovery of \Nstar{}.}
	\label{tab:obs_summary}
	\begin{tabular}{ccccccc} 
    \hline
Observation type & Telescope (instrument) & Band  & Cadence & Total integration time & Period & Notes\\
\hline
Photometry	& NGTS	&	520-890\,nm	&	12\,s & 156\,nights	&	21/09/15-03/05/16 & 21 eclipses in total\\
Photometry	& SAAO 1.0\,m (SHOC)	&	z'	& 30\,s & 16.5\,hours	&	21/09/15-25/11/16 & 1 pri., 2 sec. eclipses \\
Spectroscopy	& WHT (LIRIS)	&	zJ	& 10\,s &	60\,s	&	24/11/16-25/11/16 & ABBA nod \\
Spectroscopy	& ESO 3.6\,m (HARPS)	&	378-691\,nm & 1\,hour & 6\,hours	&	29/10/16-31/03/17 & EGGS mode\\
	\hline
    \end{tabular}
\end{table*}

\subsection{NGTS photometry}
 \Nstar\ was discovered during season 2016 of NGTS observations. It was first identified due to its red colours suggesting it was a mid-M dwarf.  NGTS operates at ESO's Paranal observatory since early 2016 and consists of an array of twelve roboticized 20~cm telescopes \citep{ngts}. The survey is optimised for detecting small planets around K and early M stars, and predicted to find $\sim$300 new exoplanets and $\sim$5600 eclipsing binaries \citep{Guenther2017a}.

\Nstar\ was observed with NGTS on 156 nights spanning over 226 days between 2015-09-21 and 2016-05-03. We obtained 216,160 images each with 10 s exposure time, covering 21 eclipses of the system. The data were reduced and analysed using the NGTS standard pipeline, described in \citet{ngts}. The pipeline performs aperture photometry using routines within CASUTools\footnote{http://casu.ast.cam.ac.uk/surveys-projects/software-release} to extract the flux for each object from all images. We use a circular aperture with a radius of three pixels. To estimate the background, a grid of background values is created which is then used to locally interpolate the background at each pixel.  To minimise systematic effects in the data, an implementation of the SysRem algorithm \citep{Tamuz2005} is then applied to the data.

The binary was detected using \textsc{orion}~\citep{ngts}, an implementation of the box-fitting least squares algorithm suggested by \citet{Kovacs2002}, although \textsc{orion} cannot fit this lightcurve accurately as it is optimised for planetary transits, not the "V" shaped eclipse of an eclipsing binary. The algorithm suggested that the most likely period is $\sim$0.87 days and the mean eclipse duration was \Nduration{} hours.

We used the centroid vetting procedure detailed in \citet{Gunther2017} to confirm there was no contamination from background objects, and that the transit seen was not a false positive. This test is able to detect shifts in the photometric centre-of-flux during transit events at the sub-milli-pixel level. It can identify blended eclipsing binaries at separations below 1\arcsec, well below the size of individual \NGTS\ pixels ($\sim$5\arcsec). We find no centroiding variation during the transit events of \Nstar.

\subsection{SAAO photometry}

We observed \Nstar\ on the SAAO 1~m telescope to attempt detection of a secondary eclipse at the NGTS suggested period. Observations were conducted in the $z'$ band using the SHOC camera \citep{coppejans13} for a combined total of 7.5 hours on the nights of 2016-10-20 and 2016-10-23 and a further 9 hours in total on the nights of 2016-11-24 and 2016-11-25. 

The data were reduced using the standard procedure with sky flats and bias frames taken during the observing run. We used the \textsc{starlink} package \textsc{autophotom} to perform the photometry of the target and comparison stars. The aperture was fixed for the data and was set to be two times the mean seeing (full width at half-maximum; \citealt{Naylor98}). This aperture size limited the impact of the background noise in the aperture. The sky background level was determined using the clipped mean of the pixel value in an annulus around the stars and the measurement errors were estimated from the sky variance. To remove atmospheric fluctuations, the light curve was divided by the light curve of one of the comparison stars.

\begin{figure}
 	\includegraphics[width=\columnwidth]{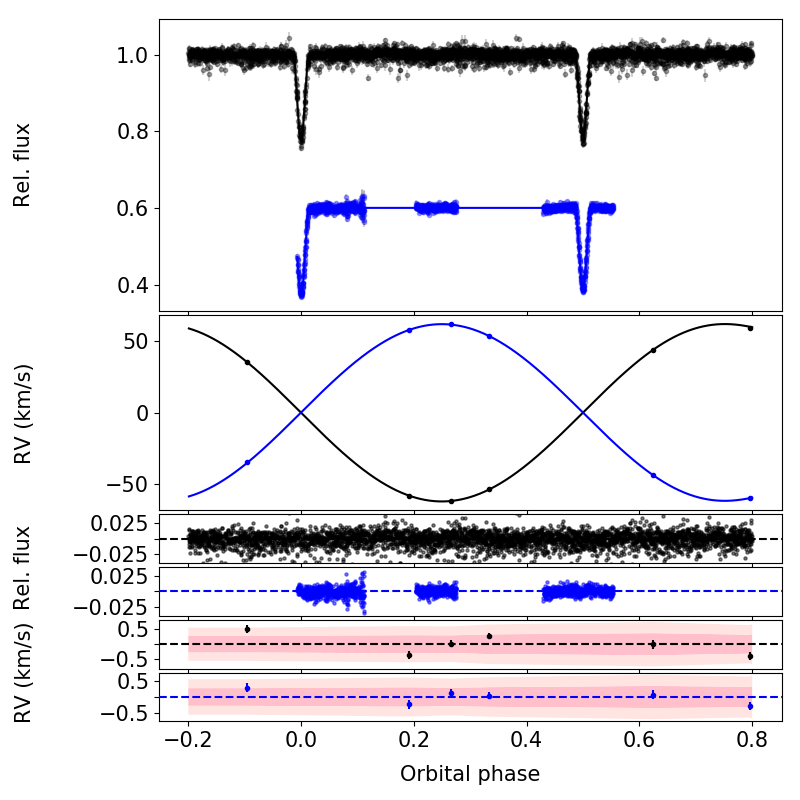}
     \caption{\textit {Top panel:} NGTS lightcurve (black) and SHOC lightcurve (blue). \textit {Second panel:} HARPS radial velocity measurements - black points and lines correspond to star A, blue corresponds to star B. \textit {Panels 3 and 4:} residuals in the top panel. \textit {Panels 4 and 5:} residuals in panel 2. All panels show the data folded on the best fitting period as determined from the global modelling. The median posterior models are plotted as solid lines; pink and beige shaded regions show the 1 and 2-sigma confidence intervals respectively. Residuals are O-C values.}
     \label{fig:data}
 \end{figure}

\begin{figure}
 	\includegraphics[width=\columnwidth]{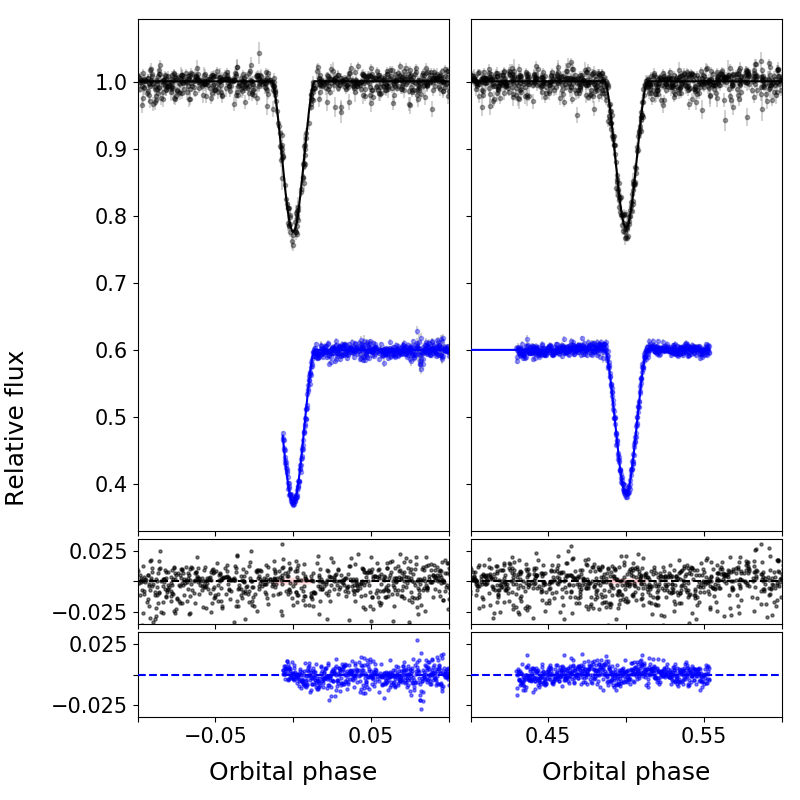}
     \caption{Focused zoom of light curves in Fig. \ref{fig:data} for clarity, depicting primary and secondary eclipses. \textit{Top panel:} NGTS lightcurve (black) and SHOC lightcurve (blue). The median posterior models are plotted as solid lines. \textit{Bottom panels:} O-C residuals. The 1 and 2-sigma confidence intervals are much smaller than the scatter of the data points.}
     \label{fig:lc_zoomed}
 \end{figure}

In addition to the data at the location of the secondary eclipse, we obtained a total of 3 eclipses of the binary over the observing period. However, we did not detect any small secondary eclipse in these data, or any reflection effect indicative of a large temperature difference between the two binary components. We thus determined that it was likely \textsc{orion} had found half the true period and that the system contains a near-equal mass binary in a $\sim$\Nperiod{} day orbit. This is confirmed by the slight difference in depths of the eclipses. The phase folded SAAO lightcurve, with the best fitting model on this period is shown in Figs.~\ref{fig:data} \& \ref{fig:lc_zoomed}, along with the NGTS equivalent.

 
\subsection{LIRIS near-IR spectrum}
We obtained a near--IR  zJ spectrum of the unresolved binary \Nstar\ on the night of 2016-10-10, using LIRIS \citep{Manchado98} on the William Herschel Telescope in service time as part of proposal SW2016b03. We observed using 10~sec exposures and an ABBA nod pattern to allow the removal of the sky background. In total we observed for 60~sec on target, with an additional observation of an A0 star, HIP24915  as a telluric standard observed immediately afterwards.  The data were reduced using the \textsc{lirisdr} \textsc{iraf} package  and telluric corrected using  the method described in \citet{vacca03}.

\begin{figure}
\begin{center}
\scalebox{0.3}{\includegraphics[]{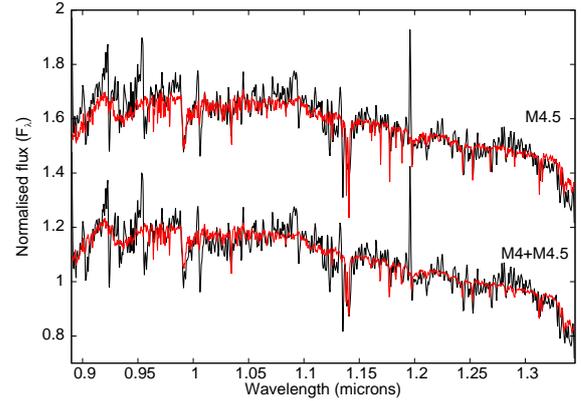}}
\caption{\label{spt} The zJ spectrum (black) of \Nstar\ and the best fit single and combined models (red).}
\end{center}
\end{figure}

We confirmed that the spectrum was of a M dwarf star. To more accurately determine the spectral type we normalised the spectrum to 1 between 1.2 and 1.3 $\mu$m,  and performed a sum of the squares of the differences to determine  our best fit spectral type using standard star M dwarf templates from the IRTF Spectral  library website. We used an M1 (HD42581), M3 (Gl388), M4 (Gl213), M4.5 (Gl268), M5 (Gl51) and M6 (Gl406) template \citep{cushing05,rayner09}. The best fit to a single star template was to spectral type M4.5. As the system is a binary, and the observations were obtained at phase 0.89, we also combined the templates. The best combined fit (RMS 0.056) was to a M4+M4.5 template, which was a better fit (RMS 0.060) than to the M4.5 template alone (Figure \ref{spt}).

\subsection{Radial velocities}
We obtained radial velocity measurements for \Nstar\ with the HARPS spectrograph \citep{harps} on the ESO 3.6\,m telescope at \LSO, Chile between 2016-10-29 and  2017-03-31.  Due to the faint optical magnitude of \Nstar\, ($g$=17.657), we used the HARPS in the high efficiency fibre link (EGGS) mode, as for NGTS-1b \citep{bayliss17}. This mode utilises a sightly larger fibre	(1.4" compared with the 1.0" in nominal mode).  This gives a higher S/N spectrum, albeit with a slightly lower spectral resolution (R=85000 compared with R=110000).  To further increase the S/N we took long time exposures (1-hour) for each epoch.  

We cross-correlated the resulting 1-D spectrum against an M5 template, and used the resulting double-peaked cross-correlation function (CCF) to determine the velocities of the two components.  Since both components of this binary are similar in spectral properties, it was not possible to distinguish the components from the CCF alone.  However given we knew precisely the phase and period from the photometric data, we could assign each peak to a given component.

The resulting radial velocities showed a variation on the period derived from the SHOC data that confirms the suggestion that this system is a near-equal mass binary. We list the radial velocity measurements in Table~\ref{tab:rvs} and plot the phase-folded radial velocities and our best fit model in Figure \ref{fig:data}. The residuals to the best fit RV model scatter well above the formal uncertainties, which is likely due to the high stellar activity on these M dwarfs.

\begin{table*}
	\centering
	\caption{Radial Velocities for \Nstar\ A and B.}
	\label{tab:rvs}
	\begin{tabular}{cccccc} 
    \hline
BJD$_\mathrm{TDB}$			&	RV		&RV error &	FWHM& 	contrast&BIS\\
(-2,450,000)	& (\kms)& (\kms)& (\kms)&(\kms) & (\kms)\\
		\hline
        HARPS  STAR A& & & & &\\
        \hline
7690.85803156	&-35.771&	0.121&	6.649&	8.4&	-0.102\\
7754.70346918	&85.604&	0.136&	8.498&	6.5&	-1.166\\
7756.64128522	&62.012&	0.144&	8.696&	6.9&	0.628\\
7837.53861917	&-32.322&	0.142&	7.449&	7.1&	4.596\\
7839.53507805	&-27.342&	0.113&	7.20&	7.6&	-0.068\\
7843.53981688	&70.205&	0.150&	9.029&	5.7&	0.0670\\

\hline
        HARPS  STAR B& & & & &\\
        \hline
7690.85803156&	88.360&	0.140&	7.863&	6.8&	0.256\\
7754.70346918&	-33.334&	0.137&	7.848& 6.5&	-0.097\\
7756.64128522	&-8.338&	0.144&	8.808&	6.3&	0.458\\
7837.53861917	&84.186&	0.140&	7.671&	6.8&	-0.059\\
7839.53507805	&80.156&	0.111&	7.569&	6.9&	-0.104\\
7843.53981688&	-17.249&	0.143&	7.613&	6.9&	0.073\\
        \hline
 
	\end{tabular}
\end{table*}

\section{Analysis}
\subsection{Activity}
We have identified a stellar flare from this system during our analysis, detected on the night of the 2015-12-31. This flare was detected during an automated search for a minimum of three consecutive points above a 6 $MAD$ level, where $MAD$ is the median absolute deviation of a night of data. It was then verified by visual inspection of all flagged flare candidates across all targets. For \Nstar\ this flare was the only flagged candidate. To calculate the energy of this flare we use the method from \citet{Shibayama13}, which assumes the flare to be a blackbody-like emitter of $9000\pm500$K. Previous works have identified such emission from M dwarf flares \citep[e.g.][]{Mochnacki80, Hawley92} and this method has consequently been used for M dwarf flares in other works \citep[e.g.][]{Yang17}. As we do not know which star the flare came from, we calculate flare energies for both the primary and secondary source. For these calculations we use the effective temperatures and stellar radii for each source given in Tab.\,\ref{system_params}. We calculate the flare energies as $4.2^{+1.1}_{-0.9}\times 10^{32}$ and $3.9^{+0.9}_{-0.7}\times 10^{32}$~erg for the primary and secondary source respectively. We can also calculate the fractional flare amplitude $\frac{\Delta F}{F}$ for each source, where $\Delta F = F_{\rm flare} - F $ with $F$ being the quiescent flux of the star. We calculate $\frac{\Delta F}{F}$ as 1.03$\pm0.11$ and 0.85$\pm0.09$ for the primary and secondary source respectively. We estimate the scale time (the time where the flare is above half of its maximum amplitude) of this flare as three minutes. To check the probability that this flare is not due to noise, we calculate the false alarm probability (FAP) according to equation 12 of \citet{Pitkin14}, assuming our noise is Gaussian. For our detection threshold and the night of the flare, we calculate a FAP to be negligible, making us confident this flare is not due to noise. To estimate the full duration we use the region where the light curve is above 1$\sigma$ from the median. From this, we estimate the full duration as five and a half minutes. This short duration, large amplitude flare is similar to those observed by \citet{Ramsay13} on the M4V star KIC 5474065, although at least double in amplitude to the largest in the Ramsay sample. HARPS spectra also show H$\alpha$ emission from \Nstar, highlighting one or both stars may be magnetically active. Magnetically active stars have been observed to flare with greater frequency than their inactive counterparts \citep[][]{Hilton11,Hawley14}. However, flares of large amplitude such as the one observed are possible from both active and inactive stars. Consequently, even if we assumed we had H$\alpha$ from a single star, it is difficult to fully correlate the H$\alpha$ emission and flare to a specific source.

\begin{figure}
\begin{center}
\scalebox{0.4}{\includegraphics[]{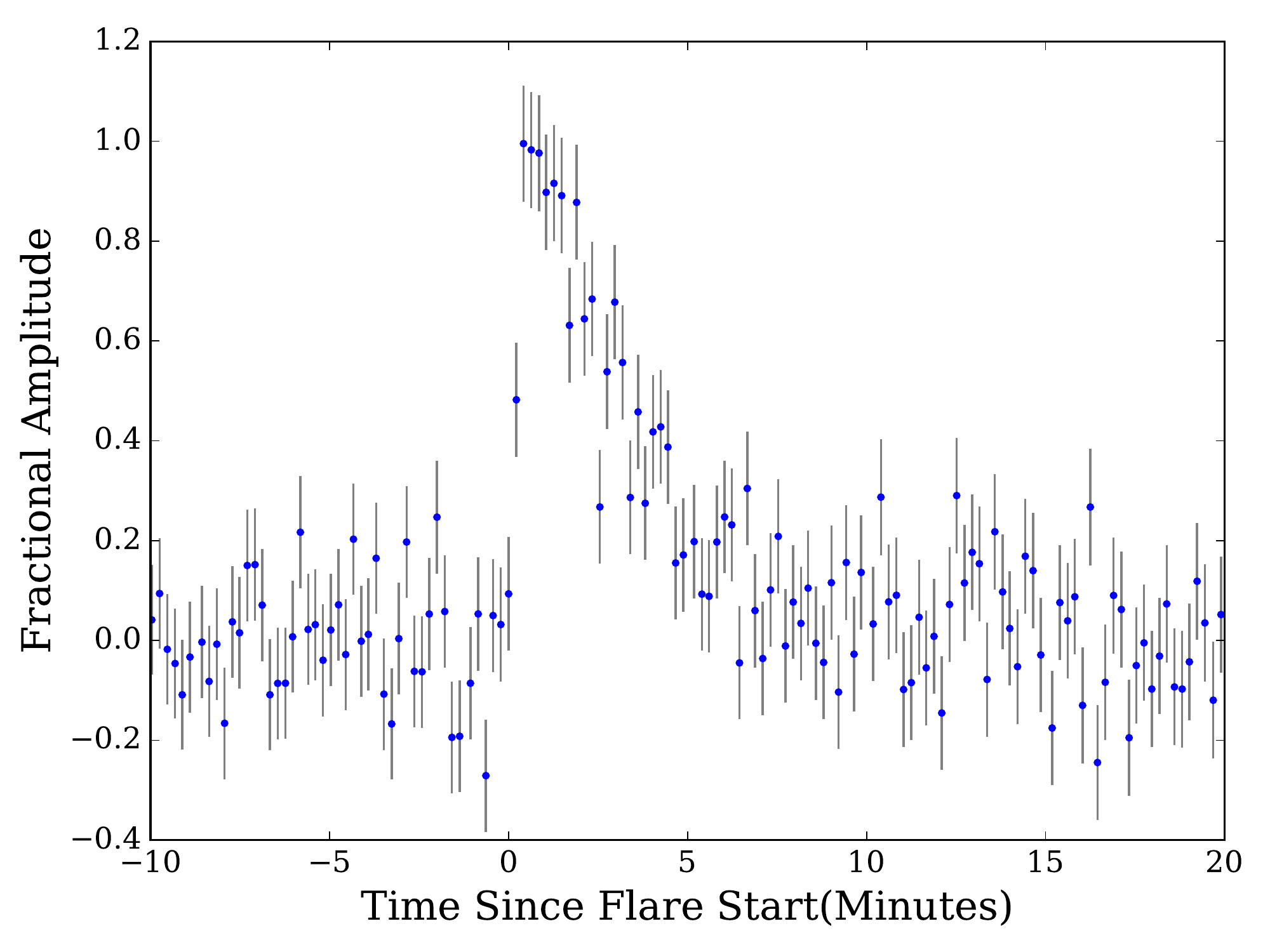}}
\caption{\label{stellar_flare} The flare observed on \Nstar\ on 2015 December 31st. The fractional flux shown here assumes the flare is from the secondary source.}
\end{center}
\end{figure}

The position of the target has been observed in the ROSAT All-Sky Survey for a total of 430~s; however, the target was not detected in X-rays. Assuming a mean coronal temperature of 5\,MK, this converts to an upper limit of $2.5 \times 10^{-13}$~erg~s$^{-1}$~cm$^{-2}$ on the X-ray flux in the 0.1--2.4\,keV energy band. \Nstar\, was also not detected in the XMM-Slew survey \citep{saxton2008}. The XMM-Slew survey data provides a model-dependent upper limit of $6.49 \times 10^{-13}$~erg~s$^{-1}$~cm$^{-2}$ in the 0.2--2\,keV band, and across the 0.2--12\,keV band the upper limit is $2.77 \times 10^{-12}$~erg~s$^{-1}$~cm$^{-2}$. These limits are at a level expected for M~dwarfs of this effective temperature, as  shown by \citet{Stelzer2013}. 

\subsection{Rotation}

As part of the determination of the system parameters we performed a detailed search for signs of stellar rotation periods in the NGTS photometry. Prior to any period search the eclipse events (both primary and secondary eclipses) were removed from the photometry. The data was also pre-whitened to remove a $\sim$1.1 day period due to a windowing of the data, and a $\sim$29.3 day period due to lunar variations which could add complexity to the output of the period search. Finally, the data was binned to 0.01 days (a necessary constraint of one of the period searching algorithms).

\begin{figure*}
 	\includegraphics{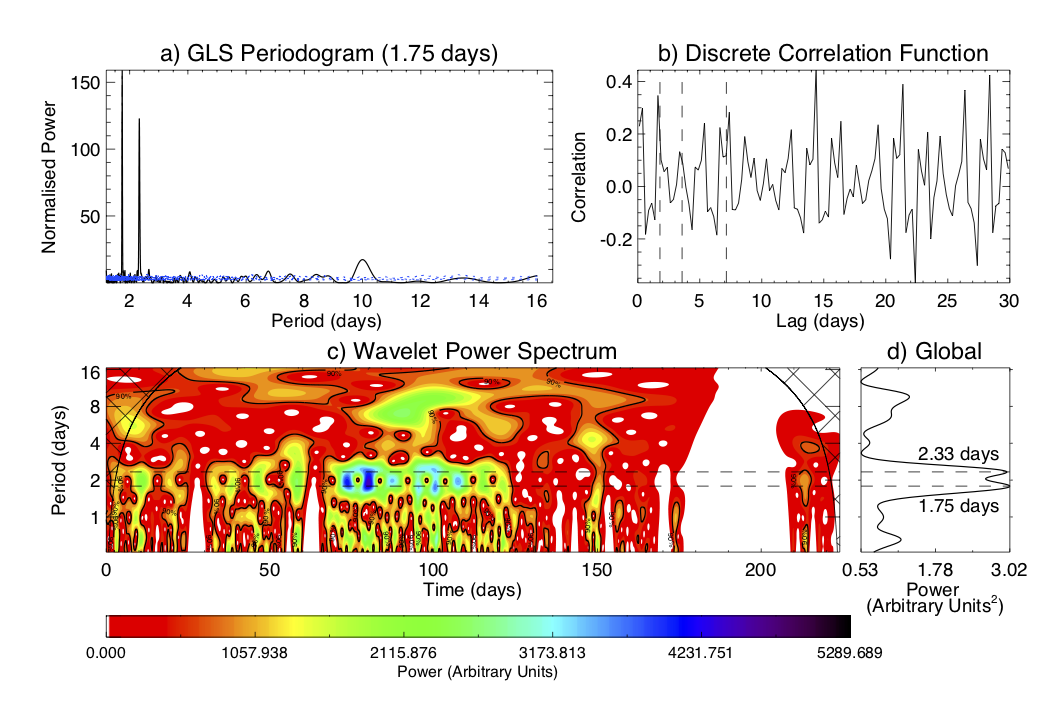}
     \caption{ a) Generalised Lomb Scargle Periodogram of \Nstar. Two strong peaks are seen at $\sim$1.75 days and $\sim$2.33 days. The dotted blue line is the 99\% FAP value. b) The Discrete Correlation Function over 30 lags, the dashed vertical lines are set at 1, 2, and 4 times the $\sim$1.75 day period seen in the GLS. c) wavelet analysis of \Nstar. 0 Days corresponds to the first measured NGTS data point. The colour-scale indicates the power contribution at each periodicity at any given time, with the black contour line indicating the 90\% confidence limit on the detection. The cross-hatched regions at the edge of the power spectrum are the `cone of influence' power inside these regions are affected by aliasing and windowing effects and therefore cannot be trusted. d) the wavelet power spectrum summed along the time direction. The same $\sim$1.75 and $\sim$2.33 day periods are seen here.}
     \label{fig:wavelet}
\end{figure*}

We used three independent methods to search for additional periodicity in the light-curve -- the Generalised Lomb Scargle (GLS) periodogram \citep{Zechmeister2009}, wavelet analysis \citep{Torrence1998}, and the Discrete Correlation Function (DCF) \citep{Edelson1988}. Figure~\ref{fig:wavelet} shows the output for these period searches, with all methods searching in the range of $1.2-16$ days. The GLS periodogram shows a strong peak at $\sim$1.74 days (which is close to the orbital period of the system) and a second strong peak at $\sim$2.33 days (see panel a of Figure~\ref{fig:wavelet}). The 99 per cent False Alarm Probability (FAP) was also calculated for the GLS periodogram using 300 bootstraps of the data in order to sample the window function (shown as the blue dotted line in panel a of Figure~\ref{fig:wavelet}). The peaks identified are well above this limit and are therefore not a result of alias effects. The DCF shows a modulation in the strength of its correlation on a similar $\sim$1.74 day period (shown as dashed lines in panel b of Figure~\ref{fig:wavelet}). The wavelet analysis also shows a similar result, with the wavelet power spectrum showing high power at periods close to those identified in the GLS, and the global wavelet spectrum (created by summing the wavelet power spectrum in time) showing peaks at the same periods seen in the GLS periodogram (panels c and d of Figure~\ref{fig:wavelet}, respectively).

\begin{figure}
 	\includegraphics[width=\columnwidth]{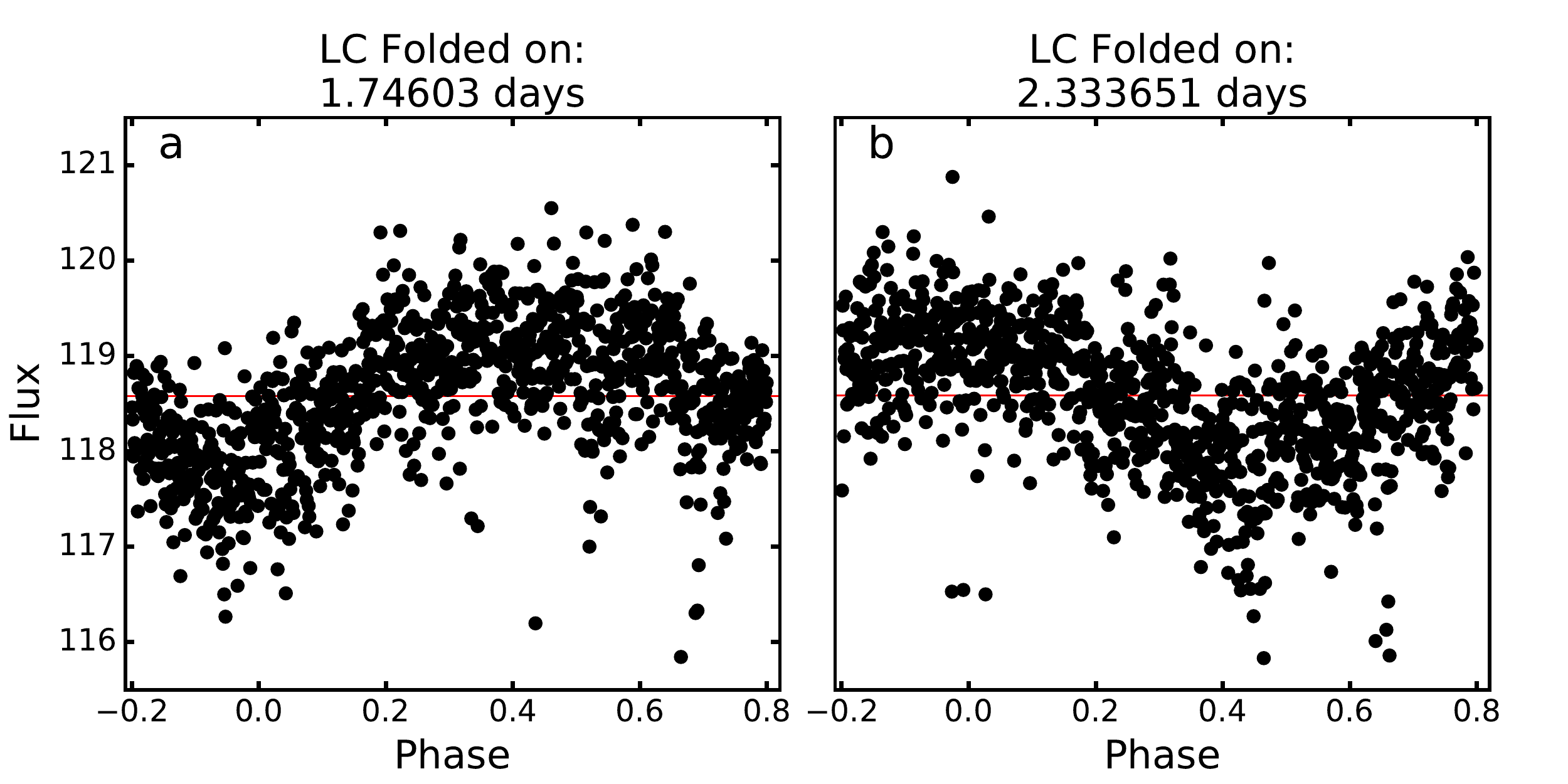}
     \caption{Phase-folded light-curve of the \Nstar, with the folding period shown on the title of the plot. a) this is close to the orbital period of the binary so implies near synchronous rotation period of one (or both) of the stars with the orbital period. b) another strong period found, at approximately $1.33 \times 1.75$ days, which could represent the rotation of one of the stars, but is more likely a harmonic effect.}
     \label{fig:LC_folded}
\end{figure}

Phase-folding the data on the $\sim$1.74 day period shows a obvious modulation of the light-curve at $\sim$2 per cent level (see Figure~\ref{fig:LC_folded}). The variation does not arise from ellipsoidal modulations due to distortions of the component stars, as this would present as two `peaks' and `troughs' in the phase-folded light-curve. In addition, the binary components lie well within their Roche lobes, and thus ellipsoidal modulation is not expected for this system. The out-of-eclipse variations seen in the phase-folded light-curve must then be attributed to activity on the surface of one (or both) of the component stars. The presence of H$\alpha$ emission and the flare event confirms that the system is magnetically active. This would imply that one (or both) of the stars has its rotation period close to the orbital period, and hence be (or be close to) tidally locked.

The light-curve was also phase-folded on the $\sim$2.33 day period, which also shows variability (though less sinusoidal) and could represent a situation where one of the stars is not tidally locked (see panel b of Figure~\ref{fig:LC_folded}). This period, however, is $\frac{4}{3}P_{rot}$ and could be the effect of a close harmonic. To try to confirm this the analysis was re-run with the $\sim$1.75 day period removed in the pre-whitening. Periods close to the 2.33 day period were found but on phase-folding no signal could be seen, indicating that a harmonic effect is indeed the most likely the source of the $\sim$2.33 day signal.

\begin{table}
	\centering
	\caption{Stellar Properties for \Nstar\, from Pan-STARRS \citep{panstarrs16}, NGTS \citep{ngts}, 2MASS \citep{2mass}, Gaia \citep{gaia} and WISE \citep{wise}.}
	\label{tab:stellar}
	\begin{tabular}{ccc} 
	Property	&	Value		&Source\\
	\hline
	Gaia I.D.		&	DR2 2957804068198875776	& Gaia \\
    R.A.		&	05:22:18.2					&		\\
	Dec			&	-25:07:09.17	&		\\
    $\mu_{\alpha}$ (\masy)& $20.215\pm0.092$  & Gaia\\
    $\mu_{\delta}$ (\masy) & $24.63\pm0.10$ & Gaia\\
    dist(pc)&	$54.41\_{-0.21}^{+0.22}$& Gaia\\
    dist (pc)&		$54.33\pm0.21$ & this work\\
    $G$			&$15.29$	&Gaia\\
	$g$			&$17.657\pm0.004$		&Pan-STARRS\\
	$r$			&$16.356\pm0.002$		&Pan-STARRS\\
	$i$			&$14.556\pm0.002$		&Pan-STARRS\\
    NGTS		&$14.65$				& this work\\
    $z$			&$13.720\pm0.003$		& Pan-STARRS\\
    $y$			&$13.270\pm0.002$		& Pan-STARRS\\
    $J$			&$11.943\pm0.025$		&2MASS	\\
   	$H$			&$11.315\pm0.026$		&2MASS	\\
	$K_{s}$			&$11.046\pm0.023$		&2MASS	\\
    $W1$			&$10.819\pm0.023$		&WISE	\\
    $W2$			&$10.631\pm0.021$		&WISE	\\
    
	\hline
	\end{tabular}
\end{table}

\subsection{Mass, radius and orbital parameters}
To constrain the stellar masses, radii and orbital parameters of the system, we globally (simultaneously) modelled the photometric and radial velocity data. To do this we utilised the binary star and exoplanet model ELLC \citep{Maxted2016}, together with the Markov Chain Monte Carlo (MCMC) sampler, EMCEE \citep{Foreman-Mackey2013}.

Both the NGTS and z' band light curves were prepared by normalising by their median out-of-eclipse (OOE) flux. The NGTS light curve exhibited OOE sinusoidal variability with a period similar to the orbital period of the system; the variability was removed by fitting a sinusoid to the OOE region. The NGTS light curve was then binned to 10 mins in time. 

We utilised a quadratic limb darkening law (\citealt{Kopal1950}), obtaining initial coefficients and uncertainties from the LDTk package \citep{Husser2013,Parviainen2015}. Based on our previously derived stellar properties, LDTk was given the following relaxed priors: $T_{\rm eff}=3000\pm200$\,K, logg$=4.5\pm0.5$\,\cmss\, and [Fe/H]$=0\pm0.2$\,dex. 

During the fit, the limb darkening coefficients (LDC) and uncertainties from LDTk were placed as Gaussian priors on the fitted LDC parameters. A Gaussian prior of $1.00\pm0.15$ was also placed on the light ratio of the stars in each band, calculated using the ratio of spectroscopic contrasts from the RV data. Zero third-light contribution was assumed since no nearby objects were identified in the photometric aperture nor when cross-matching with GSC-2, 2MASS and GAIA (DR2) catalogues. Apart from the fixed, zero third-light contribution, all parameters were free to float. The semi-major axis of the system and stellar masses were calculated using constants from \citet{Harmanec2011}.

We ran EMCEE with 192 walkers, each with 45,000 steps, initialised in a Gaussian ball around parameters which provided a good initial fit.
Likelihood values from the fitting of each dataset were combined with equal weighting and 3000 steps were discarded as burn-in. Finally, the Gelman-Rubin criterion \citep{gelman1992} was used to check chain convergence. 

As is convention e.g. \citep{gillen17},  we define the primary as the star that, when occulted, gives the deepest eclipse, and the secondary as the occulting star. We note that these adjectives do not necessarily imply that the primary star is the more massive or brighter star, as we find to be the case in this work. 

The best fit estimations for parameters are presented in Table \ref{system_params}. We obtain the following masses and radii for the primary and secondary stars: $M_{\rm pri}$=\NstarMassA $M_{\odot}$, $M_{\rm sec}$=\NstarMassB $M_{\odot}$, $R_{\rm pri}$=\NstarRadA $R_{\odot}$, $R_{\rm sec}$=\NstarRadB $R_{\odot}$.

\begin{figure}
	\includegraphics[width=\columnwidth]{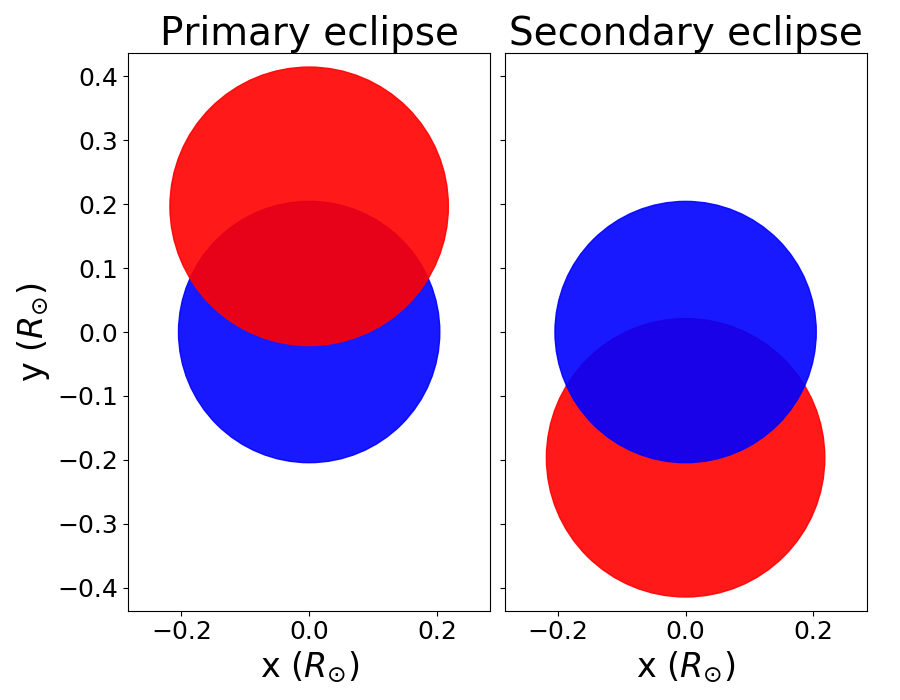}
    \caption{Diagram of system geometry for the primary (blue) and secondary (red) eclipses.}
    \label{fig:Geometry}
\end{figure}

The derived orbital inclination $i$=\OrbitalInc~degrees is expected given that the stars have grazing eclipses and the semi-major axis $a$=\SemiMajorAxis $R_{\odot}$ is consistent with a detached, short separation binary. The geometry of the system during primary and secondary eclipse is depicted in Fig. \ref{fig:Geometry}. Fig. \ref{fig:data}, shows these global fits to models for the NGTS, $z'$ band and HARPS RV datasets.

We carried out an independent check of our results by modelling the system using GP-EBOP \citep{gillen17}. The main difference between ELLC and GP-EBOP for modelling detached EBs is that GP-EBOP utilises Gaussian processes to account for the effects of stellar activity and residual systematics in the determination of the fundamental stellar parameters. In this system, the photometric and spectroscopic activity appears to be relatively small and hence the parameter estimates from both ELLC and GP-EBOP agree to within 2-sigma. Both ELLC and GP-EBOP find the secondary is slightly larger and more massive than the primary star, however GP-EBOP finds the fundamental parameters of the two stars are consistent to within their 1-sigma uncertainties and lie on a single stellar model isochrone, whereas ELLC favours a slightly inflated secondary component relative to the primary.

\begin{figure}
\includegraphics[width=\columnwidth]{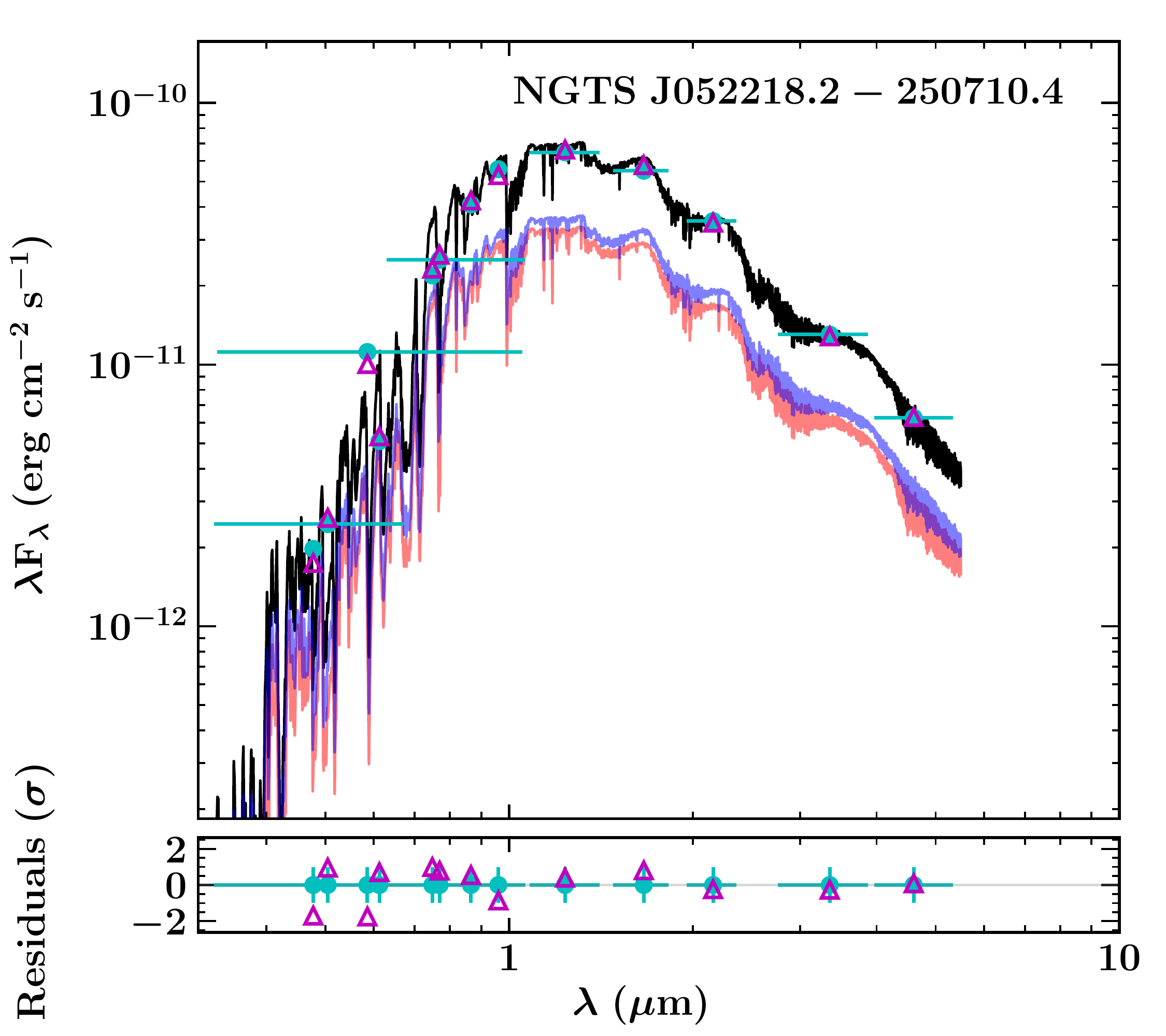}
    \caption{Spectral energy distribution (SED) of the system (cyan points) with the PHOENIX v2 model spectra fit. Top: the black line shows the combined spectrum from both stars with the magenta triangles representing the integrated spectrum in each observed bandpass. The red and blue lines indicate the contributions from the primary and secondary stars, respectively. Bottom: residuals of the fit.}
    \label{fig:SED}
\end{figure}

Following the global modelling, we determined the effective temperatures of both stars, and the distance to the system, by modelling the system's spectral energy distribution (SED) using the method presented in \citet{gillen17}. Essentially, the broadband magnitudes presented in Table~\ref{tab:stellar} are modelled as the sum of two stellar photospheres. 
We used the PHOENIX stellar atmosphere models \citep{husser13} and interpolate in $T_{\rm eff}$ -- $\log g$ space (fixing [M/H]=0) within a Markov Chain Monte Carlo (MCMC) to explore the posterior parameter space. We placed priors on the radii and surface gravities derived from our global modelling. The $T_{\rm eff}$ of both stars, and the distance and reddening to the system, all had uninformative priors. We also placed a constraint on the system parallax of $\varpi = 18.378 \pm 0.072$ from Gaia DR2, which in practice corresponds to a direct constraint on the system distance. The fit using the PHOENIX model atmospheres is shown in Figure \ref{fig:SED}.
We also tried modelling the system's SED using the BT-SETTL atmospheres \citep{allard11} but found that the BT-SETTL models overpredicted the observed blue-optical fluxes. We therefore opted to use only the PHOENIX models (rather than combining both the PHOENIX and BT-SETTL posterior distributions) for the final posterior parameters. We find temperatures of $T_{\rm pri} = 2995\,^{+85}_{-105}$\,K and $T_{\rm sec} = 2997\,^{+66}_{-101}$\,K, and a distance of $d = 54.0^{+0.2}_{-0.2}$\,pc, which is consistent with the \gaia\ distance of $54.41\_{-0.21}^{+0.22}$ \citep{gaia}. These temperatures are consistent with both stars having M5$\pm$1V spectral types, in agreement with the near-IR spectral modelling. The SED modelling results are further reported in Table \ref{system_params}.

\begin{table*}
\centering
\caption{Fitted global modelling parameters and subsequent derived parameters for the \Nstar{} system. The modal values of the posterior distributions were adopted as the most probable parameters, with the $68.3\%~(1\sigma)$ and $99.7\%~(3\sigma)$ highest probability density intervals as the error estimates.}
\label{system_params}
\renewcommand{\arraystretch}{1.4}
\begin{tabular}{|c|c|c|c|c|c|}
\hline
{\bfseries Parameter} &
{\bfseries Description} &
{\bfseries Unit} &
{\bfseries Value} & 
{\bfseries $1\sigma$ Error} & 
{\bfseries $3\sigma$ Error} \\
\hline

Fitted parameters \\ 
\cmidrule(lr){1-6}

{$\frac{R_{\rm pri}}{a}$} & radius ratio of primary to semi-major axis & none & $0.04756$ & $^{+0.00089}_{-0.00133}$ & $^{+0.0021}_{-0.0023}$ \\

$k$ & radius ratio of stars, $R_{\rm sec}/R_{\rm pri}$ & none & $1.066$ & $\pm0.047$ & $^{+0.089}_{-0.097}$ \\

$b$ & impact parameter, $\cos{(i)}/R_{\rm pri}$ & none & $0.962$ & $^{+0.017}_{-0.024}$ & $^{+0.033}_{-0.049}$ \\

$\sqrt{e} \cos \omega$ & orbital eccentricity and argument of periastron term & none & $-0.022$ & $^{+0.036}_{-0.021}$& $^{+0.067}_{-0.040}$\\

$\sqrt{e} \sin \omega$ & orbital eccentricity and argument of periastron term & none & $0.033$ & $^{+0.031}_{-0.040}$ & $^{+0.050}_{-0.080}$ \\

$P$ & orbital period & days & $1.74774445$ & $^{+0.00000036}_{-0.00000054}$ & $^{+0.00000080}_{-0.00000099}$\\

$T_{\rm c-pri}$ & epoch of primary eclipse centre & BJD & $2457288.412922$ & $^{+0.00013}_{-0.000075}$ & $^{+0.00022}_{-0.00018}$\\

{$\sigma_{\rm NGTS}$} & systematic error in NGTS light curve & norm. flux & $0.00759$ & $^{+0.00019}_{-0.00016}$ & $^{+0.00037}_{-0.00032}$ \\

{$\sigma_{z}$} & systematic error in z' light curve & norm. flux & $0.00375$ & $^{+0.00019}_{-0.00012}$ & $^{+0.00034}_{-0.00028}$ \\

{$\beta_{\rm NGTS}$} & normalised flux scale factor in NGTS data & none & $1.00117$ & $^{+0.00015}_{-0.00020}$ & $^{+0.00032}_{-0.00037}$ \\

{$\beta_{z}$} & normalised flux scale factor in z' data & none & $1.00056$ & $^{+0.00014}_{-0.00017}$ & $^{+0.00029}_{-0.00033}$ \\

$u_{\rm{NGTS-pri}}$ & linear LDC of primary in NGTS band & none & $0.469$ & $^{+0.036}_{-0.028}$ & $^{+0.067}_{-0.060}$\\

$u'_{\rm{NGTS-pri}}$ & quadratic LDC of primary in NGTS band & none & $0.280$ & $^{+0.053}_{-0.055}$ & $\pm 0.11$\\

$u_{\rm{NGTS-sec}}$ & linear LDC of secondary in NGTS band & none & $0.472$ & $^{+0.033}_{-0.030}$ & $^{+0.063}_{-0.062}$\\

$u'_{\rm{NGTS-sec}}$ & quadratic LDC of secondary in NGTS band & none & $0.278$ & $^{+0.055}_{-0.051}$ & $^{+0.11}_{-0.10}$ \\

$u_{\rm{z-pri}}$ & linear LDC of primary in z' band & none & $0.305$ & $^{+0.041}_{-0.028}$ & $^{+0.076}_{-0.062}$ \\

$u'_{\rm{z-pri}}$ & quadratic LDC of primary in z' band & none & $0.246$ & $^{+0.053}_{-0.069}$ & $^{+0.12}_{-0.13}$ \\

$u_{\rm{z-sec}}$ & linear LDC of secondary in z' band & none & $0.428$ & $^{+0.033}_{-0.035}$ & $^{+0.067}_{-0.070}$\\

$u'_{\rm{z-sec}}$ & quadratic LDC of secondary in z' band & none & $0.512$ & $^{+0.048}_{-0.073}$ & $^{+0.11}_{-0.13}$\\

$J_{\rm NGTS}$ & light ratio in NGTS band & none & $1.100$ & $^{+0.105}_{-0.099}$ & $\pm0.20$ \\

$J_{z}$ & light ratio in z' band & none & $1.051$ & $^{+0.098}_{-0.097}$ & $^{+0.20}_{-0.19}$ \\

{$K_{\rm pri}$} & radial velocity semi-amplitude of primary & \kms & $62.26$ & $^{+0.23}_{-0.18}$ & $^{+0.49}_{-0.43}$ \\

{$K_{\rm sec}$} & radial velocity semi-amplitude of secondary & \kms & $61.890$ & $^{+0.346}_{-0.060}$ & $^{+0.61}_{-0.32}$ \\

{$\Gamma_{\rm pri}$} & systemic velocity measured from primary & \kms & $26.27$ & $^{+0.16}_{-0.20}$ & $^{+0.39}_{-0.43}$ \\

{$\Gamma_{\rm sec}$} & systemic velocity measured from secondary & \kms & $26.47$ & $^{+0.17}_{-0.18}$ & $^{+0.40}_{-0.41}$ \\

{$\sigma_{\rm HARPS}$} & jitter in RV data & \kms & $0.324$ & $^{+0.163}_{-0.088}$ & $^{+0.43}_{-0.16}$ \\

\cmidrule(lr){1-6}

Derived parameters \\
\cmidrule(lr){1-6}

{$R_{\rm pri}$} & radius of primary & $R_\odot$ & $0.2045$ & $^{+0.0038}_{-0.0058}$ & $^{+0.0090}_{-0.0100}$\\

{$R_{\rm sec}$} & radius of secondary & $R_\odot$ & $0.2168$ & $^{+0.0047}_{-0.0048}$ & $^{+0.0088}_{-0.0101}$ \\

$m_{\rm pri}$ & mass of primary & $M_\odot$ & $0.17391$ & $^{+0.00153}_{-0.00099}$ & $^{+0.0032}_{-0.0027}$ \\

$m_{\rm sec}$ & mass of secondary & $M_\odot$ & $0.17418$ & $^{+0.00193}_{-0.00059}$ & $^{+0.0036}_{-0.0022}$ \\

{$a$} & semi-major axis of system & $R_\odot$ & $4.2945$ & $^{+0.0158}_{-0.0039}$ & $^{+0.029}_{-0.017}$ \\

$i$ & orbital inclination & $deg$ & $87.404$ & $^{+0.016}_{-0.019}$ & $^{+0.033}_{-0.036}$ \\

$e$ & orbital eccentricity & none & $0.00097$ & $^{+0.00214}_{-0.00097}$ & $^{+0.00977}_{-0.00097}$\\

$\log\,g_{\rm pri}$ & primary surface gravity & \cmss & $5.873$ & $^{+0.046}_{-0.048}$ & $^{+0.089}_{-0.097}$ \\

$\log\,g_{\rm sec}$ & secondary surface gravity & \cmss & $5.745$ & $^{+0.049}_{-0.038}$ & $^{+0.101}_{-0.073}$ \\

$T_{\rm 14-pri}$ & primary eclipse duration & hours & $1.1567$ & $^{+0.0038}_{-0.0055}$ & $^{+0.0084}_{-0.0101}$ \\

$T_{\rm 14-sec}$ & secondary eclipse duration & hours & $1.1567$ & $^{+0.0038}_{-0.0055}$ & $^{+0.0084}_{-0.0101}$\\

$T_{\rm eff-pri}$ & effective temperature of primary & K & $2995$ & $^{+85}_{-105}$ \\

$T_{\rm eff-sec}$ & effective temperature of secondary & K & $2997$&$^{+66}_{-101}$ \\

$d$& distance to system &pc&$54.40$&$^{+0.20}_{-0.20}$\\

\hline
\end{tabular}
\end{table*}

\section{Discussion}

Our effective temperatures for both system components of \Nstar\ are  consistent with the values presented in Table~12 of \citet{Boyajian2012}. Our temperatures are consistent with their values for an M5.5 dwarf, however, our radii fall between those given for M3.5 and M4 dwarfs, although it should be noted that at these later spectral types, only one object was used to define each spectral bin in their table.  \citet{Boyajian2012} also note in their work that when comparing single stars to eclipsing binaries, the data from the eclipsing binaries have larger radii than single stars of the same temperature, or cooler temperatures for the same radius. This is certainly the case here, where our binary components have a radius that is larger than predicted by the  \citet{Boyajian2012} relation. This discrepancy has been suggested to be related to the presence of greater activity in binaries than in lone stars \citep{lopez07}.  Our HARPS spectra show H$\alpha$ emission from one or both components of the binary, hence may account for this discrepancy.

While there are few objects within the same parameter space to use as comparisons to \Nstar\, there are two known late near-equal mass M dwarf--M dwarf binaries that have precise masses and radii  known to a few percent with similar periods -- CM~Dra, and KOI126.  CM~Dra is a nearby ($d=14.5$~pc) eclipsing binary consisting of  two active  M4.5 dwarfs in a 1.7 day orbit with masses of  0.23  and 0. 21~M$_{\odot}$ and radii of 0.25 and 0.24~R$_{\odot}$ respectively \citep{Morales09}.  KOI126 is a triple system comprising two M dwarfs with masses of 0.24 and 0.21~M$_{\odot}$ and radii of 0.25 and 0.23~R$_{\odot}$ in a 1.767 day orbit, similar to CM~Dra. This binary is then on a much longer eccentric orbit about a third star \citep{carter11}. Both of these systems have estimated ages of 4~Gyr.  HATS551-027, is an additional system with a much longer orbital period ($\sim$4.1~days) that was discovered by \citet{Zhou2015}. Although this binary has components that are not near equal mass; M$_{1}$=0.244~M$_{\odot}$, R$_{1}$=0.261~R$_{\odot}$, M$_{2}$=0.179~M$_{\odot}$ , R$_{2}$ = 0.218~R$_{\odot}$, they do fall within the same mass-radius parameter space as the previous two systems and \Nstar.  \citet{Zhou2015} determine the secondary star in HATS551-027 is inflated by $\sim$9 per cent, whereas the primary star has a radius that agrees with models. They also detect H$\alpha$ emission from these stars indicating they may be active.

 \Nstar\ has individual stellar masses of below 0.3~M$_{\odot}$, and hence the stars are expected to be fully convective (e.g. \citealt{spada}), as is the case for CM Dra, KOI126 and  HATS551-027. However, while \Nstar\ has a similar period and spectral types to CM Dra and KOI126, the M dwarf masses are much lower at 0.17391$^{+0.00153}_{-0.00099}$ and 0.17418$^{+0.00193}_{−0.00099}$~M$_{\odot}$, meaning not only is it the lowest mass field age eclipsing M dwarf binary known to date, is also a valuable addition to the mass--radius relation for M~dwarfs.
 
There are additional systems known such as AD2615 which is a young (age $\sim$ 800 Myr) system reported to be a member of the Praesepe open star cluster \citep{gillen17}, JW 380, which is a pre-main sequence binary of age 2-3 Myr with masses 0.26$\pm$0.02 and 0.15$\pm$0.01 M$_{\odot}$ \citep{irwin07} and 2MASS02405152+5245066, a 250 Myr F star+ low mass M dwarf (0.188$\pm$ 0.014 M$_{\odot}$) system \citep{eigmul}. However, these young systems are expected to have faster rotation and more activity than older, field age systems such as \Nstar\ and so we limit our discussion to field age M dwarfs only.

\begin{figure*}
\begin{center}
\scalebox{0.6}{\includegraphics[trim={0 7cm 0 0},clip]{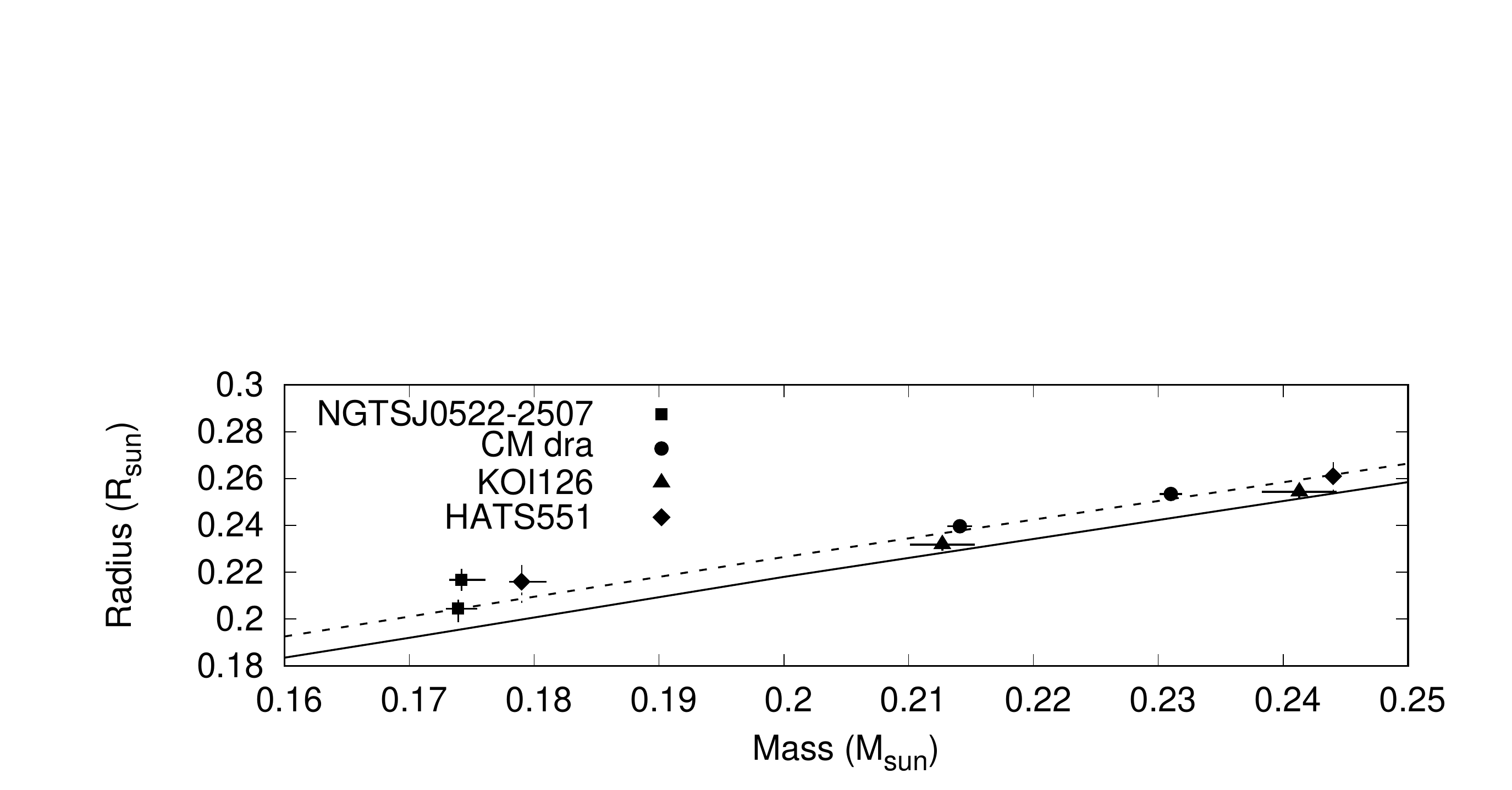}}
\scalebox{0.6}{\includegraphics[trim={0 5cm 0 5cm},clip]{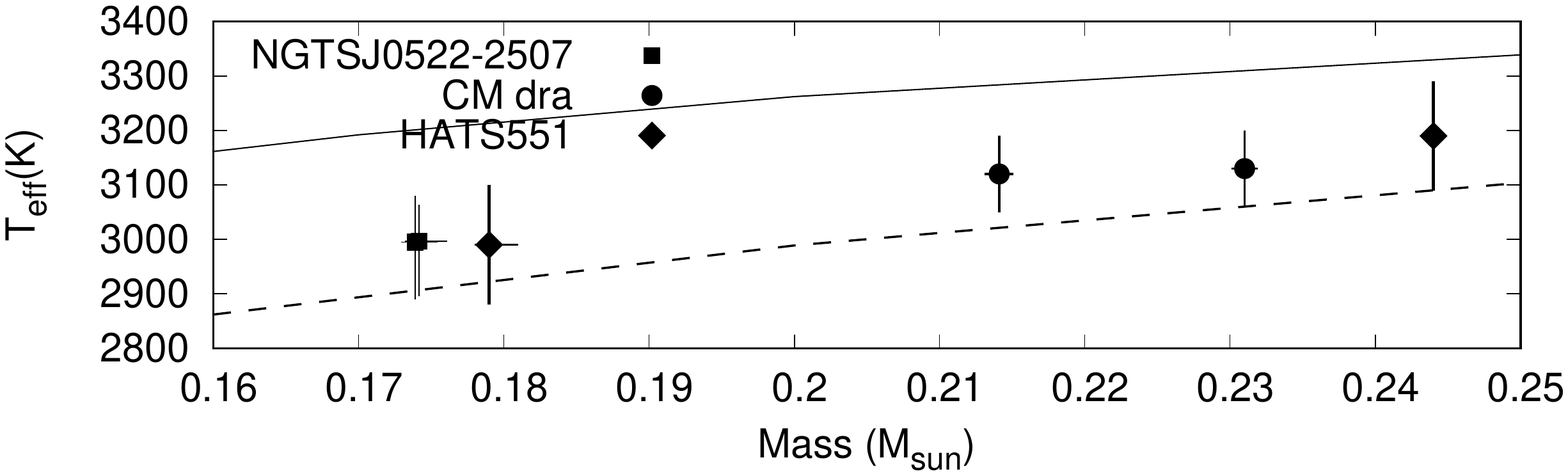}}
\caption{\label{MRteff} Mass-radius and Mass-T$_{\rm eff}$ relations for the four known eclipsing low mass binaries. \Nstar\ is plotted with squares, CM Dra with circles, HATS551-027 with  diamonds and KOI126 with triangles. KOI126 is not present in lower panel as there are no T$_{\rm eff}$ measurements for the binary in the literature. The isochrones from \citet{baraffe15} (solid line) and the PARSEC isochrones (\citealt{parsec}: dotted line) are also plotted with ages 5 Gyr.}
\end{center}
\end{figure*}

These four late-M dwarf binaries, \Nstar\ , CM Dra, HATS551 and KOI126, are shown to all fall on the model isochrones for 5 Gyr in mass-radius space (Figure \ref{MRteff}). A precursor to these isochrones were also used in \citet{Demory2009}, and our results are consistent with theirs on single stars within this mass and radius regime. It should however be noted that \citet{Demory2009} only have stellar parameters for a M0.5 and an M5.5 dwarf, so we cannot make any further comparison here without data for the spectral types in between.

These results are consistent with those of \citet{Zhou2015}. Figure \ref{MRteff} shows the four field age M dwarf binaries compared to the Baraffe \citep{baraffe15} and PARSEC isochrones \citep{parsec}, the latter of which have been calibrated for low mass stars by \citet{chen14}. The binaries fall between the isochrones in the T$_{\rm eff}$ vs mass relation, with all objects nearer to, but hotter than the PARSEC isochrones, and cooler than the Baraffe ones. In the mass-radius relation. KOI126 seems to fit better with the Baraffe isochrones than the PARSEC models, but this is the only system that shows this, with the remaining three sitting closer to the PARSEC models. This discrepancy may again be related to activity, as KOI126 is not reported to to be active, in the way the three other systems have. Indeed, \citet{feiden12} suggest that measurements of the system made to date may have been taken during a period of inactivity. 
\Nstar\ is cooler than the Baraffe models suggest it should be, but this is a feature that is consistent with all the binaries presented here. The secondary star component of \Nstar\ is larger than both models would suggest, which may indicate it is indeed active, particularly when combined with the fact one component is flaring, and there are out of transit modulations to the lightcurve.  The primary star sits on the PARSEC model.

\section{Conclusions}

We have discovered a new eclipsing, late M dwarf binary from NGTS. This is a near-equal mass binary, comprising two M4-5 dwarfs in a $\sim$1.7 day orbit. We determine the masses of the M dwarfs with uncertainties below 1 per cent and their radii with uncertainties below 3 per cent. In this mass regime, there are very few well characterised M dwarfs, thus highlighting the importance of this system. The masses of these stars are lower than the $\sim$0.2 M$_{\odot}$ of both CM Dra and KOI126, both of which have similar periods.  Comparisons to isochrones show that the components of  \Nstar\ have  masses and radii consistent with an age of 5 Gyr although the secondary star in \Nstar\ does show a larger radius than is predicted by models,  potentially due to activity.

\section*{Acknowledgements}

This publication is based on data collected under the NGTS project at the ESO La Silla Paranal Observatory. The NGTS instrument and operations are funded by the consortium institutes and by the UK Science and Technology Facilities Council (STFC; project reference ST/M001962/1). The WHT and its service programme are operated on the island of La Palma by the Isaac Newton Group of Telescopes in the Spanish Observatorio del Roque de los Muchachos of the Instituto de Astrofisica de Canarias. This paper also uses observations made at the South African Astronomical Observatory (SAAO).  SLC acknowledges support from LISEO at the University of Leicester. EG acknowledges support from the Winton foundation. MRG and MRB are supported by an STFC consolidated grant (ST/N000757/1). PJW, RGW and TL are supported by an STFC consolidated grant (ST/P000495/1). JSJ acknowledges support by FONDECYT grant 1161218 and partial support by CATA-Basal (PB06, CONICYT). This work utilises the ELLC exoplanet and binary star model developed by P.F.L. Maxted and we thank him for his specific recommendations for using this model.




\bibliographystyle{mnras}
\bibliography{mdwarf}

\bsp	
\label{lastpage}
\end{document}